\documentclass[preprint,showpacs,preprintnumbers,amsmath,amssymb]{revtex4}

\usepackage{graphicx}
\usepackage{dcolumn}
\usepackage{bm}

\begin{document}

\vspace{5mm}

\newcommand{\goo}{\,\raisebox{-.5ex}{$\stackrel{>}{\scriptstyle\sim}$}\,}
\newcommand{\loo}{\,\raisebox{-.5ex}{$\stackrel{<}{\scriptstyle\sim}$}\,}

\title{Mechanisms for production of hypernuclei beyond the neutron and proton 
drip lines}

\author{N.~Buyukcizmeci,$^{1,2}$ A.S.~Botvina,$^{2,3,4}$ 
J.~Pochodzalla,$^{4,5}$ M.~Bleicher$^{2}$}

\affiliation{$^1$Department of Physics, Selcuk University, 42079 Kampus, 
Konya, Turkey} 
\affiliation{$^2$Frankfurt Institute for Advanced Studies, J. W. Goethe 
University, D-60438 Frankfurt am Main, Germany} 
\affiliation{$^3$Institute for Nuclear 
Research, Russian Academy of Sciences, 117312 Moscow, Russia} 
\affiliation{$^4$Helmholtz-Institut Mainz, J. Gutenberg-Universit{\"a}t, 
D-55099 Mainz, Germany} 
\affiliation{$^5$Institut f{\"u}r Kernphysik and PRISMA Cluster of 
Excellence, J. Gutenberg-Universit{\"a}t, D-55099 Mainz, Germany}

\date{\today}

\begin{abstract}

We analyze hypernuclei coming from fragmentation and multifragmentation 
of spectator residues obtained in relativistic ion collisions. These hypernuclei 
have a broad distribution in masses and isospin. They reach beyond the 
neutron and proton drip lines, and they are expected to be stable 
with respect to neutron and proton emission. This gives us the opportunity 
to investigate the properties of exotic hypernuclei, as well as 
the properties of normal nuclei beyond the drip lines, which can be 
produced after weak decay of such hypernuclei. 

\end{abstract}

\pacs{21.80.+a, 25.70.Mn, 25.70.Pq, 24.60.-k}

\maketitle

\section{Introduction}

In high-energy nuclear reactions strange hadrons (baryons and 
mesons) are produced abundantly, and they are strongly involved in 
the reaction process. When strange baryons (hyperons) are captured 
by nuclei, hypernuclei 
are formed, which can live long enough in comparison with nuclear 
reaction times. The detailed study of hypernuclei is essential 
to overcome the very limited experimental possibilities 
of exploring hyperon-nucleon ($YN$) and hyperon-hyperon ($YY$) 
interactions in elementary scattering experiments 
($Y=\Lambda,\Sigma,\Xi,\Omega$). Double- and multi-strange nuclei 
are especially interesting, because they can provide information about 
the hyperon--hyperon interaction and strange matter properties. 
Studying hypernuclei helps us to understand the structure of conventional 
nuclei too \cite{japan}, and to extend the nuclear chart into 
the strangeness sector \cite{cgreiner,greiner}. 
Strangeness is an important conception for the construction of 
theoretical models of strong interactions \cite{schramm}. It is also 
important that hyperons should be abundantly produced in nuclear 
matter at high densities, which are realized in the core of neutron stars 
\cite{schaffner}. To describe these astrophysical conditions realistically it is necessary to study all aspects of hyperon interactions in the 
laboratory and select theoretical models which pass careful 
comparison with experimental data. 

Traditionally, information about hypernuclei is obtained from 
spectroscopy \cite{japan}. 
The theoretical studies are mainly concentrated on calculating the 
structure of nearly cold hypernuclei with a baryon density around the nuclear 
saturation density, $\rho_0 \approx 0.15$ fm$^{-3}$. However, 
a quite limited set of reactions was generally used for producing 
hypernuclei: Reactions with the production of few particles, including kaons, 
are quite effective for 
triggering single hypernuclei, and by using kaon beams one can produce 
double hypernuclei. The goal of this paper is to demonstrate that one can 
essentially extend the frame of hypernuclear studies and produce exotic 
hypernuclei if new many-body reactions are involved. 

One should remember that hyperons were discovered in the 1950s in reactions 
of nuclear multifragmentation induced by cosmic rays \cite{danysz}. 
During the last 20 years of research great progress has been made in the investigation 
of multifragmentation reactions, mainly associated with heavy-ion 
collisions (see, e.g., \cite{smm,aladin97,EOS,ogul} and references therein). 
This gives us the opportunity to apply a well-known theoretical method 
adopted for description of these reactions for production of hypernuclei too 
\cite{bot-poch,dasgupta}. On the other hand, it was noticed long ago 
that the absorption of hyperons in spectator regions 
of peripheral relativistic ion collisions is a promising way to produce 
hypernuclei \cite{wakai1,cassing,giessen,botvina2011}. Also, central 
collisions of relativistic heavy ions can lead to the production of 
light hypernuclei \cite{Steinheimer}. Corresponding experimental 
evidence has been reported \cite{Nie76,Avr88}. Recent 
sophisticated experiments 
have confirmed observations of hypernuclei in such reactions, in both  
peripheral \cite{saito-new} and central collisions \cite{star}. 

Here we pay special attention to the formation of hypernuclei in the spectator 
region of peripheral relativistic ion collisions. Current research 
concerns mainly light hypernuclei produced in reactions with light 
projectiles \cite{saito-new,botvina2012}. However, there is a promising 
opportunity to study the production of large and exotic hypernuclei coming 
from reactions with large projectiles and targets \cite{botvina2011,bot2012}. 
In particular, multifragmentation decay of excited hyperspectator matter 
is a perspective mechanism \cite{bot-poch,dasgupta}. Below we investigate 
how exotic (neutron-rich and proton-rich) hypernuclei can be obtained 
in future experiments.


\section{Formation and disintegration of hyperon-rich spectator matter}


A detailed picture of peripheral relativistic heavy-ion collisions has 
been established in many experimental and theoretical studies. 
Nucleons from overlapping parts of the projectile and target 
(participant zone) interact strongly with themselves 
and with other hadrons produced in primary and secondary collisions. 
Nucleons from nonoverlapping parts do not interact intensively, and 
they form residual nuclear systems, 
which we call spectators. In all transport models 
the production of hyperons is associated with nucleon-nucleon collisions, 
e.g.,  p+n$\rightarrow$n+$\Lambda$+K$^{+}$, or collisions of secondary 
mesons with nucleons, e.g., $\pi^{+}$+n$\rightarrow \Lambda$+K$^{+}$. 
Strange particles can be produced in the participant zone, however, 
particles can rescatter and undergo secondary interactions. As a result 
the hyperons produced populate the whole momentum space around the 
colliding nuclei, including the vicinity of nuclear spectators. 
Such hyperons can be absorbed by the spectators if their kinetic energy 
(in the rest frame of the spectator) is lower than the potential generated 
by neighboring spectator nucleons.  
The process of formation of spectator hypermatter was investigated in 
Ref. \cite{botvina2011} within transport approaches: the Dubna cascade model 
(DCM) and ultra-relativistic quantum molecular dynamics (UrQMD) model. 
It was concluded that already at a beam energy of 2 A GeV single and double 
hyperspectators can be noticeably produced, while at an energy of a few 
tens of giga-electron volts per nucleon the formation of multistrange hyperspectators becomes feasible. 

In the following these excited spectators disintegrate in normal nuclei 
and nuclei containing hyperons. As our calculations and analyses 
of experimental data obtained in similar heavy-ion reactions 
\cite{smm,aladin97,EOS,ogul} show, the residual spectator nuclei have rather high 
excitation energies, and consequently, they should undergo multifragmentation 
with a characteristic time of about 100 fm/$c$. 
Generalization of the statistical multifragmentation model 
(SMM) \cite{smm} into the strangeness sector by including $\Lambda$ hyperons 
has been done in Ref.~\cite{bot-poch}. Here we summarize it. 

The model assumes that a hot nuclear spectator with total 
mass (baryon) number $A_0$, charge $Z_0$, number of $\Lambda$ 
hyperons $H_0$, and temperature $T$ expands to a low density 
freeze-out volume, where the system is in chemical equilibrium. The 
statistical ensemble includes all breakup channels composed of 
nucleons and excited fragments with mass number $A$, charge $Z$, and 
number of $\Lambda$'s $H$. The primary fragments are formed in 
the freeze-out volume $V$. We use the excluded volume approximation 
$V=V_0+V_f$, where $V_0=A_0/\rho_0$ ($\rho_0\approx$0.15 fm$^{-3}$ 
is the normal nuclear density), and parametrize the free volume 
$V_f=\kappa V_0$, with $\kappa \approx 2$ \cite{aladin97,EOS,ogul}. 

Nuclear clusters in the freeze-out volume are described as follows: 
Light fragments with mass number $A < 4$ are treated as elementary
particles with corresponding spin and translational degrees of
freedom ("nuclear gas"). Their binding energies were taken from
experimental data \cite{japan,smm,Bando}. Fragments with $A=4$
are also treated as gas particles with table masses, however, some
excitation energy is allowed $E_{x}=AT^{2}/\varepsilon_0$
($\varepsilon_0 \approx$16 MeV is the inverse volume level density
parameter \cite{smm}), which reflects the presence of excited states
in $^{4}$He, $^{4}_{\Lambda}$H, and $^{4}_{\Lambda}$He nuclei.
Fragments with $A > 4$
are treated as heated liquid drops. In this way one can study the
nuclear liquid-gas coexistence of hypermatter in the freeze-out
volume. The internal free energies of these fragments are
parametrized as the sum of the bulk ($F_{A}^B$), the surface
($F_{A}^S$), the symmetry ($F_{AZH}^{\rm sym}$), the Coulomb
($F_{AZ}^C$), and the hyper ($F_{AH}^{\rm hyp}$) energies:
\begin{equation} \label{freenergy}
F_{AZH}(T,V)=F_{A}^B+F_{A}^S+F_{AZH}^{\rm sym}+F_{AZ}^C+F_{AH}^{\rm hyp}~~.
\end{equation}
The first three terms
are written in the standard liquid-drop form \cite{smm}:
\begin{equation} \label{bulk}
F_{A}^B(T)=\left(-w_0-\frac{T^2}{\varepsilon_0}\right)A,
\end{equation}
\begin{equation} \label{surface}
F_{A}^S(T)=\beta_0\left(\frac{T_c^2-T^2}{T_c^2+T^2}\right)^{5/4}A^{2/3},
\end{equation}
\begin{equation} \label{symmetry}
F_{AZH}^{\rm sym}=\gamma \frac{(A-H-2Z)^2}{A-H}.
\end{equation}
The model parameters $w_0=16$ MeV, $\beta_0=18$ MeV, $T_c=18$ MeV
and $\gamma=25$ MeV were extracted from nuclear phenomenology and
provide a good description of multifragmentation data
\cite{smm,aladin97,EOS,ogul}. The Coulomb interaction of the
fragments is described within the Wigner-Seitz approximation, and
$F_{AZ}^C$ is taken as in Ref.~\cite{smm}.

The new term is the free hyperenergy $F_{AH}^{\rm hyp}$. We assume
that it does not change with temperature, i.e., it is determined
solely by the binding energy of the hyperfragments. 
We have suggested the liquid-drop hyperenergy term \cite{bot-poch} 
\begin{equation} \label{hyp}
F_{AH}^{\rm hyp}=(H/A)\cdot(-10.68 A + 21.27 A^{2/3}).
\end{equation}
In this formula the binding energy is proportional to the fraction
of hyperons in the system ($H/A$). The second part represents the
volume contribution reduced by the surface term and thus resembles a
liquid-drop parametrization based on the saturation of the nuclear
interaction. The linear dependence at a low $H/A$ is in agreement
with theoretical predictions \cite{greiner} for hypermatter. 

The breakup channels are generated according to their 
statistical weight. In the grand canonics this leads to the following 
average yields of individual fragments: 
\begin{eqnarray} \label{yazh} 
Y_{\rm AZH}=g_{\rm AZH}\cdot V_f\frac{A^{3/2}}{\lambda_T^3} 
{\rm exp}\left[-\frac{1}{T}\left(F_{AZH}-\mu_{AZH}\right)\right], 
\nonumber\\ 
\mu_{AZH}=A\mu+Z\nu+H\xi~, 
\end{eqnarray} 
Here $g_{\rm AZH}$ is the ground-state degeneracy factor of species 
$(A,Z,H)$, $\lambda_T=\left(2\pi\hbar^2/m_NT\right)^{1/2}$ is the 
nucleon thermal wavelength, and $m_N$ is the average 
nucleon mass. The chemical potentials $\mu$, $\nu$, and $\xi$ are 
responsible for the mass (baryon) number, charge, and strangeness 
conservation in the system. They can be found from the balance 
equations: 
\begin{eqnarray} 
\sum_{AZH}A Y_{\rm AZH}=A_0, 
\sum_{AZH}Z Y_{\rm AZH}=Z_0, 
\sum_{AZH}H Y_{\rm AZH}=H_0. 
\nonumber 
\end{eqnarray} 

Previously we have demonstrated within this model \cite{bot-poch} 
that the fragment mass distributions are 
quite different for fragments with different strangeness contents. This means
that the multifragmentation of excited hypernuclear systems
proceeds in a different way compared with conventional
nuclei. The reason is the additional binding energy of hyperons in nuclear 
matter. It was also shown  that the yields of fragments with two $\Lambda$'s
depend essentially on the binding energy formulas (i.e., on details of 
$\Lambda N$ and $\Lambda \Lambda$ interactions) used for the calculations 
\cite{bot-poch,samanta}. Therefore, an analysis of double hypernuclei can 
help to improve these mass formulas and reveal information about 
the hyperon-hyperon interaction. In Ref.~\cite{lorente} the 
decay of light excited hypersystems was considered within the framework 
of the Fermi breakup model. It was also concluded that the 
production rate of single and double hypernuclei is directly related to 
their binding energy. In this work we extend our analysis to systems 
containing up to four hyperons, which may be produced during the dynamical 
stage of relativistic heavy-ion collisions \cite{botvina2011,bot2012}.

\section{Neutron and proton drip lines of normal nuclei and hypernuclei}

\subsection{Hypernucleus binding}

The treatment of hyperfragments in the freeze-out volume is important 
in the SMM. As discussed, we have suggested an approach 
motivated by the successful application of the liquid-drop approximation 
describing multifragmentation of normal nuclei \cite{smm,aladin97,EOS,ogul}. 
On the other hand, this approximation suits well for practical using in the 
developed model. We should note that the properties of the primary
fragments  may change in the medium compared to the vacuum owing to
the proximity of other fragments (see, for example, 
Ref.~\cite{Typel}). Presently, there is evidence
that the symmetry energy \cite{ogul,LeFevre,Iglio,Souliotis} and surface
energy \cite{Botvina06} of hot fragments in multifragmentation may
be modified. 
Because the binding energies of hypernuclei are mostly not
known even in vacuum, this problem is naturally included in
searching for a reliable mass formula for hypernuclei within this
approach. 

However, it is important to demonstrate the qualitative consistence of 
the approach at temperature $T=$0 with the known experimental 
data and with more sophisticated calculations. Presently, only a 
few tens of masses of single hypernuclei (mostly light ones) have been 
experimentally established \cite{japan,Bando}, and only 
very limited information on double hypernuclei is available. In Fig.~1 
we show the experimental data on the separation energy of $\Lambda$ 
hyperons in hypernuclei, together with our liquid-drop approximation 
[Eq.~(\ref{freenergy})] 
and with results of relativistic mean-field (RMF) calculations in 
Refs.~\cite{cgreiner} and \cite{Schaffner94}. 

\begin{figure}[tbh]
\includegraphics[width=0.8\textwidth]{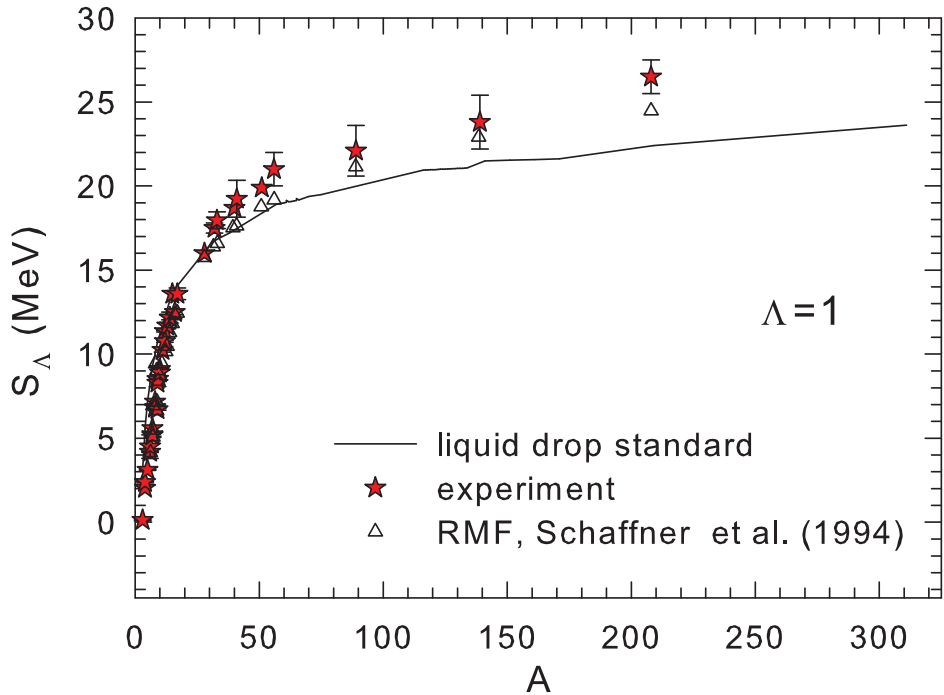}
\caption{\small{ (Color online)
Separation energy of $\Lambda$ hyperons in hypernuclei versus mass 
number. The solid line is our liquid-drop approximation at $T=0$. Stars are experimental data taken from Refs.~\cite{japan} and \cite{Bando}. Triangles are 
results of RMF calculations \cite{cgreiner,Schaffner94}. 
}}
\label{fig1}
\end{figure}

We see a reasonable agreement and reproduction of the main trend: increasing 
and saturation of separation energy of hyperons (i.e., their binding 
energy) with mass number of nuclei. Therefore, the liquid-drop approximation 
can be used for estimation of single hypernuclei. In Fig.~2 we compare 
the predictions of the RMF model with our approach for some multiple hypernuclei. 
There is a similar agreement in binding energies and reproduction of the 
trend of increasing binding energy with increasing hyperon number. 

\begin{figure}[tbh]
\includegraphics[width=0.8\textwidth]{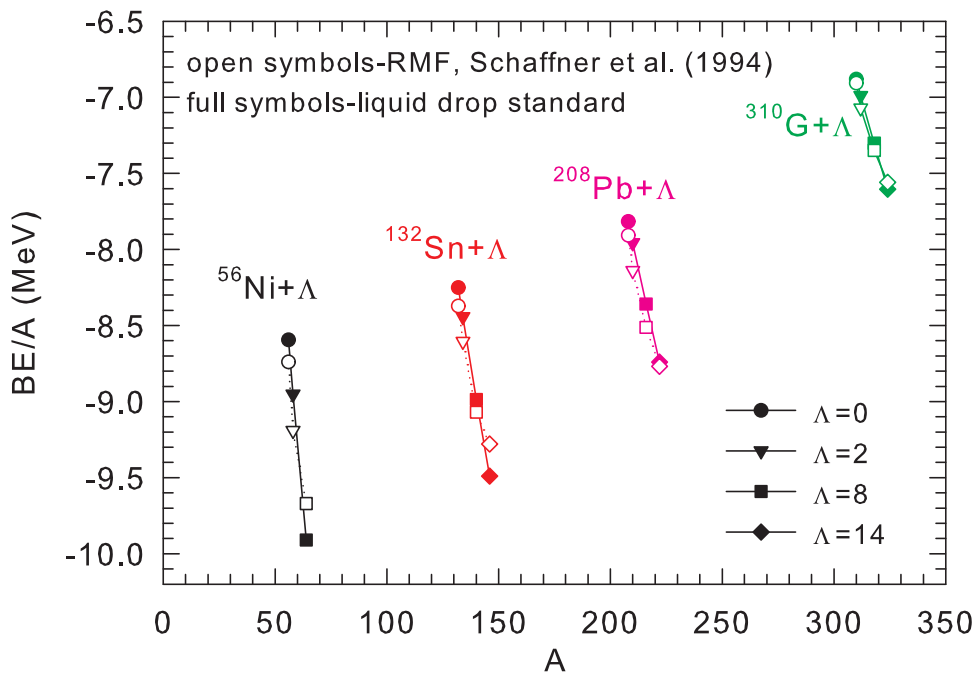}
\caption{\small{ (Color online)
Binding energies (per baryon) of Ni, Sn, Pb, and G (Z=126) $\Lambda$ 
hypernuclei 
depending on the number of $\Lambda$ hyperons. Open symbols are RMF 
calculations 
from Refs.~\cite{cgreiner} and \cite{Schaffner94}, filled symbols, the liquid-drop 
approximation. 
Numbers of hyperons considered are shown.}} 
\label{fig2}
\end{figure}

Moreover, from Figs.~1 and 2 we see that for intermediate-mass and heavy 
nuclei the predictions of the 
liquid-drop approximation slightly underestimate hyperon binding, 
by comparison with both data and RMF. In this case one can say that 
this approximation provides a low limit for the binding energy in both 
single and multiple hypernuclei. Note that we consider a number of 
$\Lambda$'s that is much smaller than the total number of baryons. It is 
sufficient for our purpose to determine the reaction processes leading 
to formation of exotic and multiple hypernuclei. If the baryonic fraction of 
hyperons in such nuclei is more than 10--20\% , a saturation effect in 
the binding energy for $\Lambda$ hypernuclei may take place and they may 
be converted into $\Sigma$ and other hyperons \cite{cgreiner}. However, 
a very high strangeness content is beyond the scope of this work.

\subsection{Separation energies of neutrons and protons around drip lines}

Neutron and proton drip lines are important for investigating 
the structure of nuclei. They also play a major role in 
nucleosynthesis of elements in space. Before moving to hypernuclei 
it is instructive to recall the behavior of the binding energy in the 
vicinity of drip lines in normal nuclei. Traditionally, this 
behavior is evaluated from the separation energies of neutrons and 
protons. For our purpose it is more convenient to use the separation 
energies of two neutrons and two protons, in order to avoid the 
well-known pairing correlations: 
\begin{eqnarray} 
S_{2n}=BE(A,Z)-BE(A-2,Z), \hspace{5mm}
S_{2p}=BE(A,Z)-BE(A-2,Z-2),
\end{eqnarray} 
where $BE(A,Z)$ is the nucleus binding energy. 

In Fig.~3 we show the two-neutron separation energies for fluorine 
(Z=9), molybdenum (Z=42), and lead (Z=82) 
over a wide range of neutron numbers. We have plotted 
the available experimental data taken from \cite{audi} together with 
some sophisticated calculations: RMF \cite{Yilmaz11}, 
the finite-range liquid-drop model (FRDM) \cite{Moller97}, 
the infinite nuclear matter model (INM) \cite{Nayak12}, 
a model with Skyrme forces (SkM) \cite{Chabanat98}, 
the Wigner-Kirkwood mean-field approximation (WK) \cite{Nazarewicz94}, 
the standard averaging method in the mean-field approximation (SAM) 
\cite{Nazarewicz94}, and 
Skyrme Hartree-Fock (SKF) \cite{Brown02}. 
These models were specially constructed to describe nuclear structure 
and, unfortunately, cannot be directly used in a model of nuclear 
reactions. We have also shown the results of our liquid-drop approximation, 
obtained from Eq.~(\ref{freenergy}), 
at $T=0$. One can see that it is sufficient for qualitative description of 
the data and the main trends. The discrepancies come from the shell 
structure of cold isolated nuclei. On the other hand, in multifragmentation reactions 
we expect to deal with hot nuclei in the surroundings of other nuclear species. 
As reported previously these structure effects should be washed out at 
high temperatures ($T > 1-2$ MeV) \cite{Ignatiuk,K-H-Schmidt}. It is 
interesting that some calculations (e.g., the RMF in Ref.~\cite{Yilmaz11} 
for Mo) predict structure effects, like islands of stability, at large 
numbers of neutrons N. 

\begin{figure}[tbh]
\includegraphics[width=0.6\textwidth]{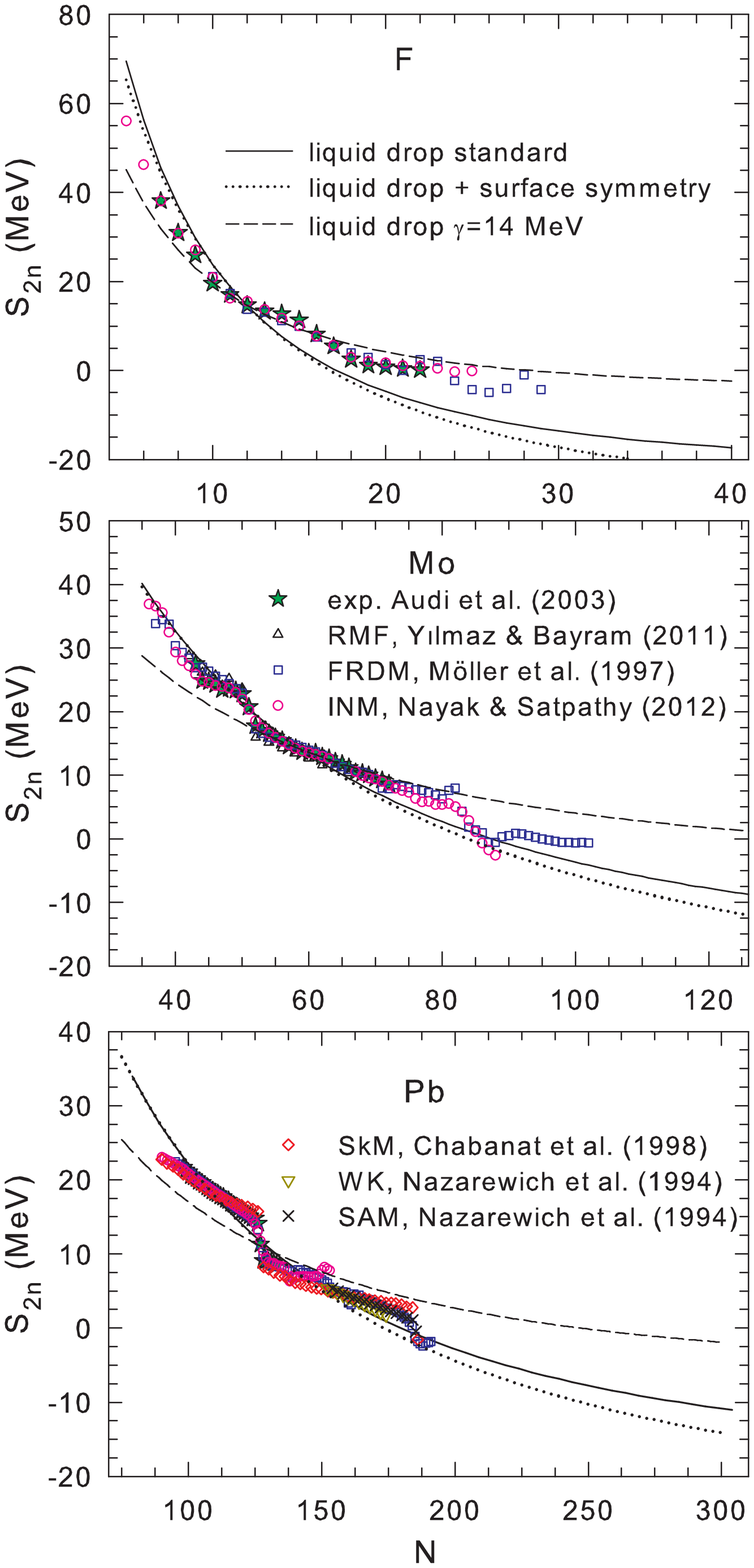}
\caption{\small{ (Color online)
Separation energy of two neutrons in fluorine (Z=9), 
molybdenum (Z=42), and lead (Z=82) nuclei 
versus neutron number. 
Stars are experimental data \cite{audi}; symbols are model 
calculations (see also the text). Lines are the liquid-drop approximation 
used in the SMM: solid line, standard symmetry energy coefficient 
($\gamma = 25$ MeV); 
dashed line, reduced symmetry energy ($\gamma = 14$ MeV); dotted line, 
including the surface symmetry energy contribution (see the text). 
}}
\label{fig3}
\end{figure}

We have investigated how modifications of the liquid-drop parameters 
may change our results. In all panels in Fig.~3 we demonstrate 
effects of a modification of the symmetry term 
[Eq.~(\ref{symmetry})], which is crucially important for these separation 
energies. Instead of the standard parametrization 
$F^{\rm sym}=\gamma (A-2Z)^2/A$, we have divided it into the volume and 
surface symmetry energy contributions, as suggested in some works 
\cite{myers}. Our aim is to compare the obtained trends, therefore, we 
have normalized the parametrizations at 
$^{21}$F, $^{102}$Mo, and $^{209}$Pb, respectively.  
As a result we found that the symmetry energy coefficient $\gamma$ 
increases with mass number; 
for example, for molybdenum $F_{m}^{\rm sym}=(34-42/A^{1/3}) (A-2Z)^2/A$. 
One can see from Fig.~3 that for all elements this division may slightly 
improve the description of small nuclei, however, it makes the description worse 
for large nuclei in the vicinity of the drip line. 
We conclude that inclusion of the surface symmetry energy into 
consideration does not help to explain the observed behavior of the 
neutron separation energy around the neutron drip line. On the other hand, 
if we simply decrease the $\gamma$ coefficient in Eq.~(\ref{symmetry}) 
from 25 to 14 MeV we obtain a trend that is less steep with N and that 
is more consistent with the experimental trend for neutron-rich isotopes. 
It is interesting that this kind of decrease in symmetry energy 
was obtained in recent analyses of experimental data on 
multifragmentation \cite{ogul,Iglio,Souliotis}. In this respect, one may 
speculate that the halo neutrons of such exotic nuclei interact with the 
core nucleons in a universal way, which looks like interactions between 
nuclear species taking place at dilute densities of the freeze-out in 
multifragmentation reactions. 

\begin{figure}[tbh]
\includegraphics[width=0.6\textwidth]{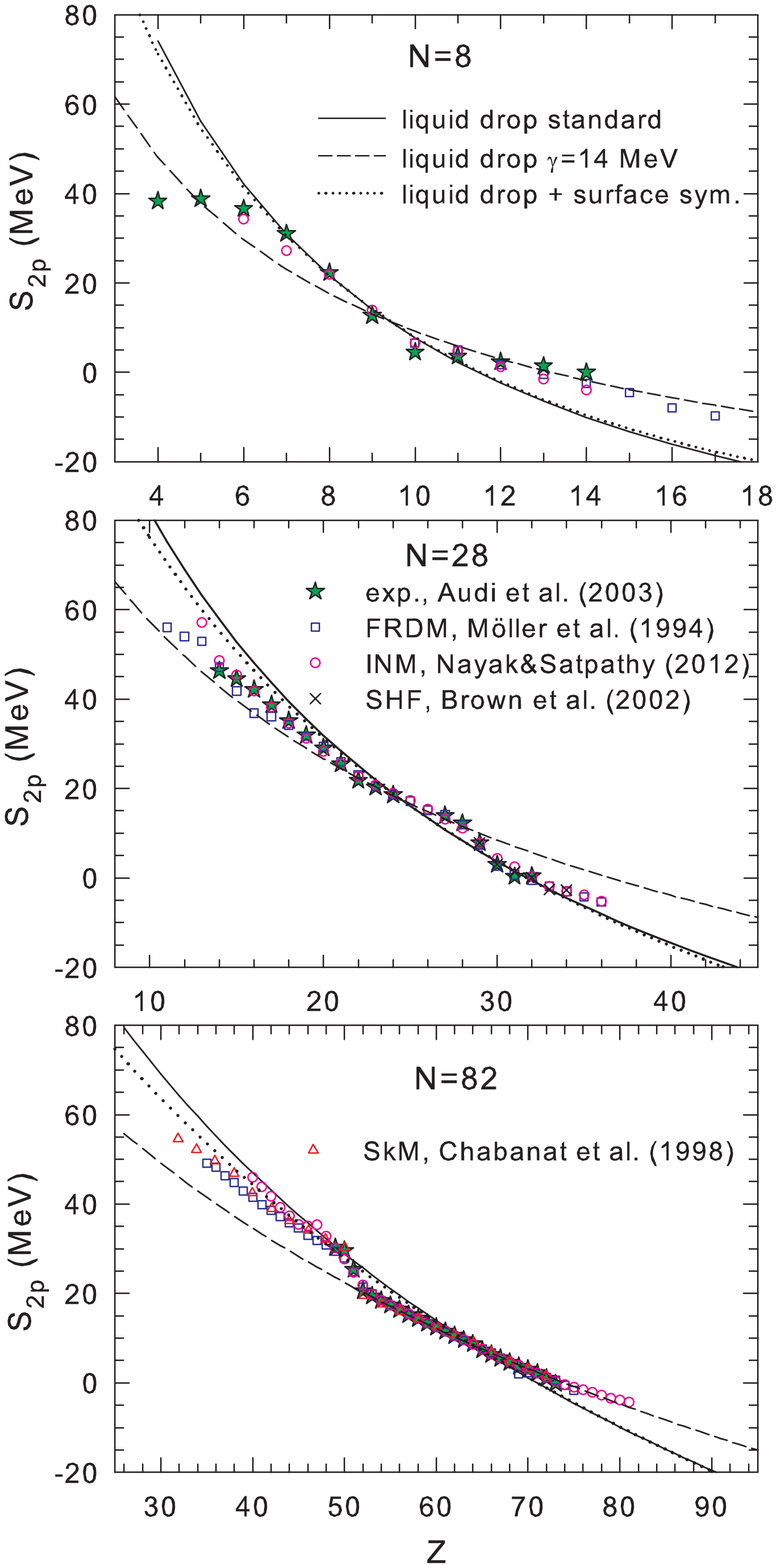}
\caption{\small{ (Color online)
Separation energy of two protons in nuclei containing 8, 28, and 82 neutrons 
versus proton number. 
Stars are experimental data \cite{audi}; symbols are model 
calculations (see also the text). Lines are 
liquid-drop approximations used in the SMM, as in Fig.~3 (see the text). 
}}
\label{fig4}
\end{figure}

By analogy, in Fig.~4 we show the two-proton separation energy for isotones 
with numbers of neutrons equal to 8, 28, and 82. 
One can see the qualitative consistence of our approximation with 
experimental data and various nuclear structure calculations.  
By examining the influence of modifications of the symmetry energy on this 
separation energy, the parametrizations were normalized at $Z$=9, 24, and 
60, respectively. The obtained trends for all masses are similar to those presented in Fig.~3. Therefore, the main conclusions drawn 
from analysis of Fig.~3 are valid for this case too.

\subsection{Influence of hyperons on separation energies in nuclei}

The presence of hyperons inside nuclei increases their binding energies, 
because of hyperons coupling with nucleons; see Eq.~(\ref{hyp}). 
This influences all structure characteristics of nuclei and is manifested 
in nuclear fragmentation 
reactions too \cite{bot-poch}. The neutron and proton separation 
energies will also be higher. We demonstrate these energies versus the 
number of neutrons and protons in nuclei in Figs.~5 and 6. 
Calculations were performed as previously with the liquid-drop 
approximation [Eq.~(\ref{freenergy})] at $T=$0, for nuclei with different 
numbers of hyperons. 

\begin{figure}[tbh]
\includegraphics[width=0.6\textwidth]{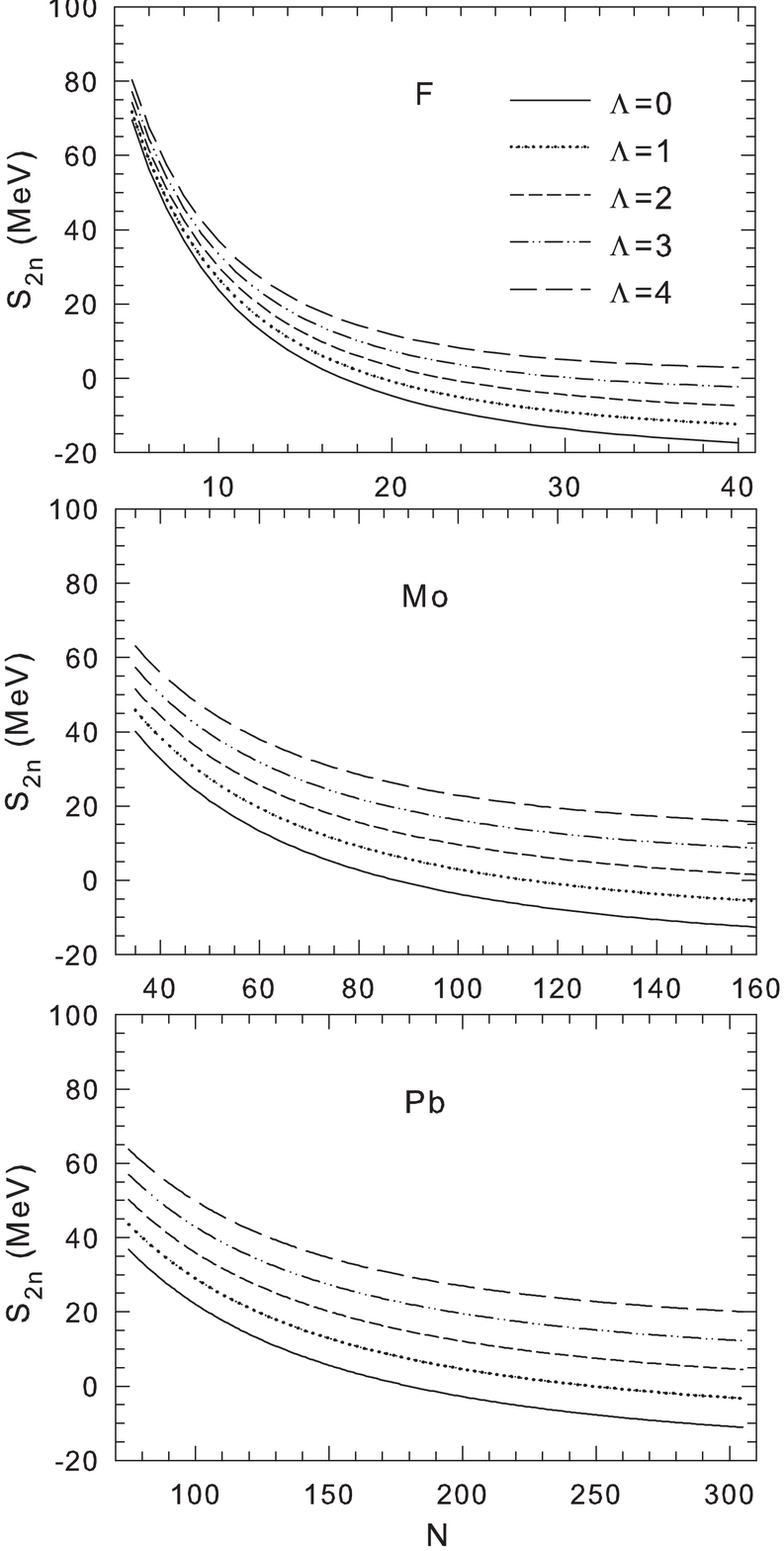}
\caption{\small{Separation energy of two neutrons in hypernuclei of fluorine, 
molybdenum, and lead, versus neutron number. Lines give different 
numbers of $\Lambda$ hyperons inside nuclei, according to 
the liquid-drop approximation (see the text). 
}}
\label{fig5}
\end{figure}

Examining the neutron drip line region around $S_{2n}=0$, in Fig.~5, 
we see that by adding only one $\Lambda$ hyperon into a Mo nucleus we shift 
its drip line to the right side by nearly 30 mass units. Very neutron-rich 
hypernuclei are possible and 
their weak decay time scale is long compared to the typical strong decay 
times. This effect is present in both small and heavy nuclei, 
however, in heavy ones it is more pronounced. Double and multiple 
hypernuclei are even more bound, therefore, their drip lines are shifted far 
more to the neutron-rich side. Such a high neutron richness can hardly be 
obtained in any reaction involving normal nuclei. In this respect, 
the production of hypernuclei may provide a natural way for accumulation of neutrons in nuclei, which can be realized in special processes. 

\begin{figure}[tbh]
\includegraphics[width=0.6\textwidth]{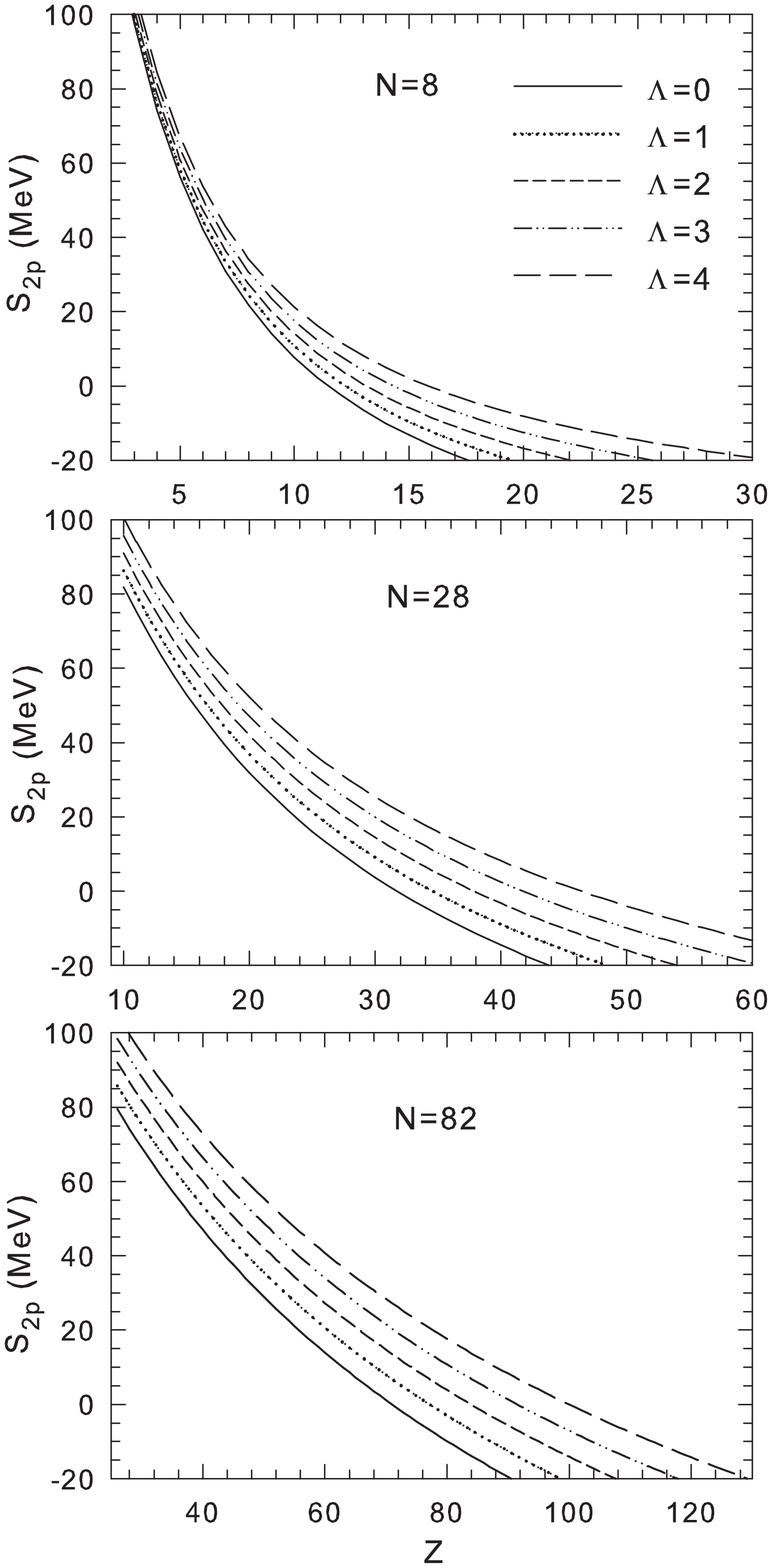}
\caption{\small{Separation energy of two protons in hypernuclei containing 8, 28, and 82 
neutrons versus proton number. Lines give different 
numbers of $\Lambda$ hyperons inside nuclei, according to 
the liquid-drop approximation (see the text). 
}}
\label{fig6}
\end{figure}

Analysis of the proton separation energies (Fig.~6) leads to similar 
conclusions: With an increasing number of hyperons the proton drip lines 
shift considerably to the side of proton richness. Because of 
Coulomb interaction this shift is less pronounced than in the neutron case. 
However, it is still a prominent effect, which could be important 
for nuclear structure studies and synthesis of new elements. 

We expect that the uncertainties in 
the hyperon-nucleon binding and the symmetry energy discussed in previous subsections will have only a minor 
influence on the results, and they will not change our 
predictions qualitatively. The reason 
is that the hyperon-nucleon interaction is a well-established fundamental concept in both theory and experiment. Recent experimental 
observations of very neutron-rich hypernuclei, e.g., the $^{6}_{\Lambda}$H 
hypernucleus \cite{6LH}, are consistent with the predicted trend. 

\section{Production of hypernuclei in fragmentation and multifragmentation 
processes} 

We analyze the problem of how these neutron-rich and proton-rich hypernuclei can 
be produced in relativistic ion collisions. It has been investigated for normal nuclei during the last 10--20 years (see, e.g., review \cite{aumann}). 
Presently there are extensive experimental projects aimed at this study, for 
example, the Fragment Separator (FRS) at GSI-Darmstadt, and one is planned 
for the future FAIR facility too \cite{frs}. 

The main physics idea is the following: Relativistic projectile ions 
undergo fragmentation/multifragmentation, spallation, and fission processes 
in interactions with targets. These processes lead to the production of a broad 
variety of exotic residues that already can be close to the neutron (proton) 
drip lines. Some of these nuclei can live long enough, in particular, 
because of their large $\gamma$ factors. 
We suggest the use of these relativistic exotic nuclei for 
new interactions with targets for producing hypernuclei. Mechanisms of such 
interactions were investigated within transport models \cite{botvina2011}. 
After the dynamical stage of the reaction excited hyperspectator residues 
are produced, which have practically the same average ratio of neutrons to 
protons as in projectiles. However, great fluctuation of 
their masses and isospin content is possible. The final hypernuclei will 
be produced after disintegration of these excited systems. They will have 
wide distributions and the hyperon contribution to binding such nuclei will make it possible to obtain many nuclei beyond the traditional drip lines. 

Below, this mechanism is illustrated with SMM calculations. As input 
for the calculations we assume dynamically produced hyperspectators 
with various masses, isospin contents, temperatures and numbers of absorbed 
hyperons. As demonstrated in Ref.~\cite{botvina2011}, at a beam energy 
of about 20 A GeV, heavy spectators can absorb up to three $\Lambda$ hyperons with 
a sizable probability, and absorption of larger numbers of hyperons is 
feasible. We consider masses of hypernuclei systems that can be naturally 
produced after the dynamical stage ($A_0=$50, 100, and 200) and with different 
isospins. Some calculations at small numbers of absorbed 
hyperons (up to $H_0=2$) were reported in Ref.~\cite{bot-poch}. Here, in order to 
generalize the results for hypermatter, in Figs.~7 and 8 we demonstrate 
results of disintegration calculations for systems with four absorbed 
hyperons. The temperature range ($T=3-5$ MeV) was adopted in order to 
investigate the region of coexistence of big and small fragments, typical 
for liquid-gas-type phase transition in finite systems, which is also observed 
in multifragmentation reactions \cite{smm}. We would like to give examples 
of systems close to the neutron drip line: For this 
reason, besides systems with an isospin content of typical large, stable 
nuclei ($Z_0/A_0 = 0.4$; Fig.~7), we present neutron-enriched systems 
that, as we conclude after examining Ref.~\cite{aumann}, may be obtained after 
reactions with beams separated by the FRS at FAIR ($Z_0/A_0 = 0.35$; Fig.~8). 

\begin{figure}[tbh]
\includegraphics[width=0.8\textwidth]{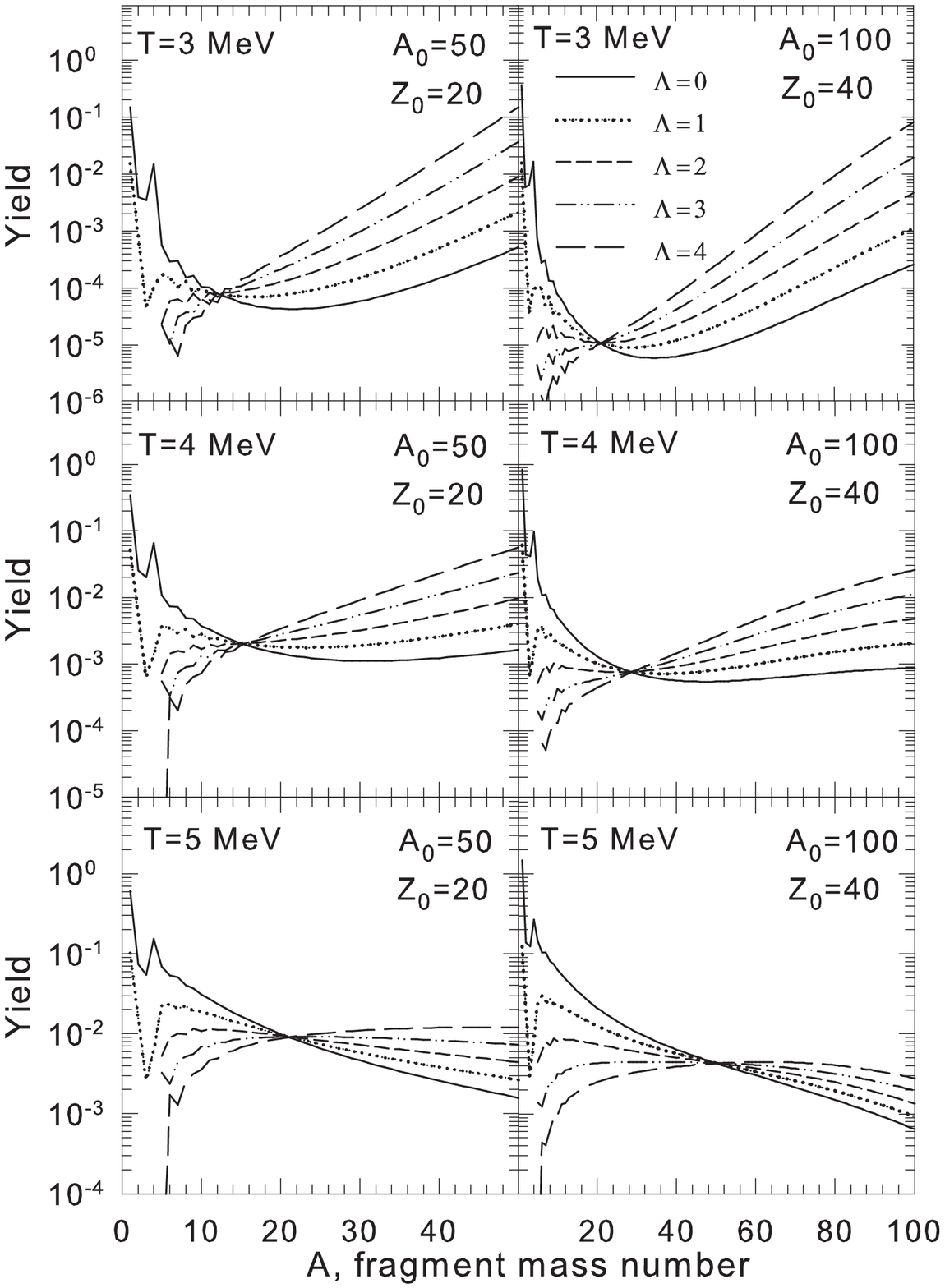}
\caption{\small{SMM predictions of yields of fragments and hyperfragments versus their 
mass number, after disintegration of excited systems containing four $\Lambda$ 
hyperons. Initial mass numbers $A_0$, charges $Z_0$, and temperatures $T$ 
of the systems are shown. Lines are calculations for 
fragments with a certain number of $\Lambda$ hyperons. Yields are given per one disintegration event. 
}}
\label{fig7}
\end{figure}

One can see a general evolution of the mass distributions with 
temperature: At a low temperature ($T=$3 MeV) we have a "U-shape" distribution 
that consists of two maxima, at the lightest (nucleons and light clusters) 
and largest (close to the system size) fragments, and the "valley" 
in between. The yield of intermediate-mass fragments increases with increasing 
temperature and at about $T=5$ MeV we obtain a "plateau"-like distribution. At higher temperatures we will have an exponential decrease in yield with 
mass number \cite{bot-poch}. This picture was solidly established in 
multifragmentation reactions with normal nuclei \cite{smm,EOS,ogul}. However, 
the presence of hyperons causes interesting consequences: 
At moderate temperatures hyperons are 
predominantly accumulated in big fragments because of the high binding 
energy. At low temperatures, when the largest nuclei survive, one can be 
sure that they contain practically all the $\Lambda$'s of the system. This 
main result will not change if one considers a canonical statistical 
ensemble for description of the system's disintegration 
(see Ref.~\cite{dasgupta}), 
though some minor details will be different from the adopted grand canonics 
of Eq.~(\ref{yazh}). 
 
\begin{figure}[tbh]
\includegraphics[width=0.8\textwidth]{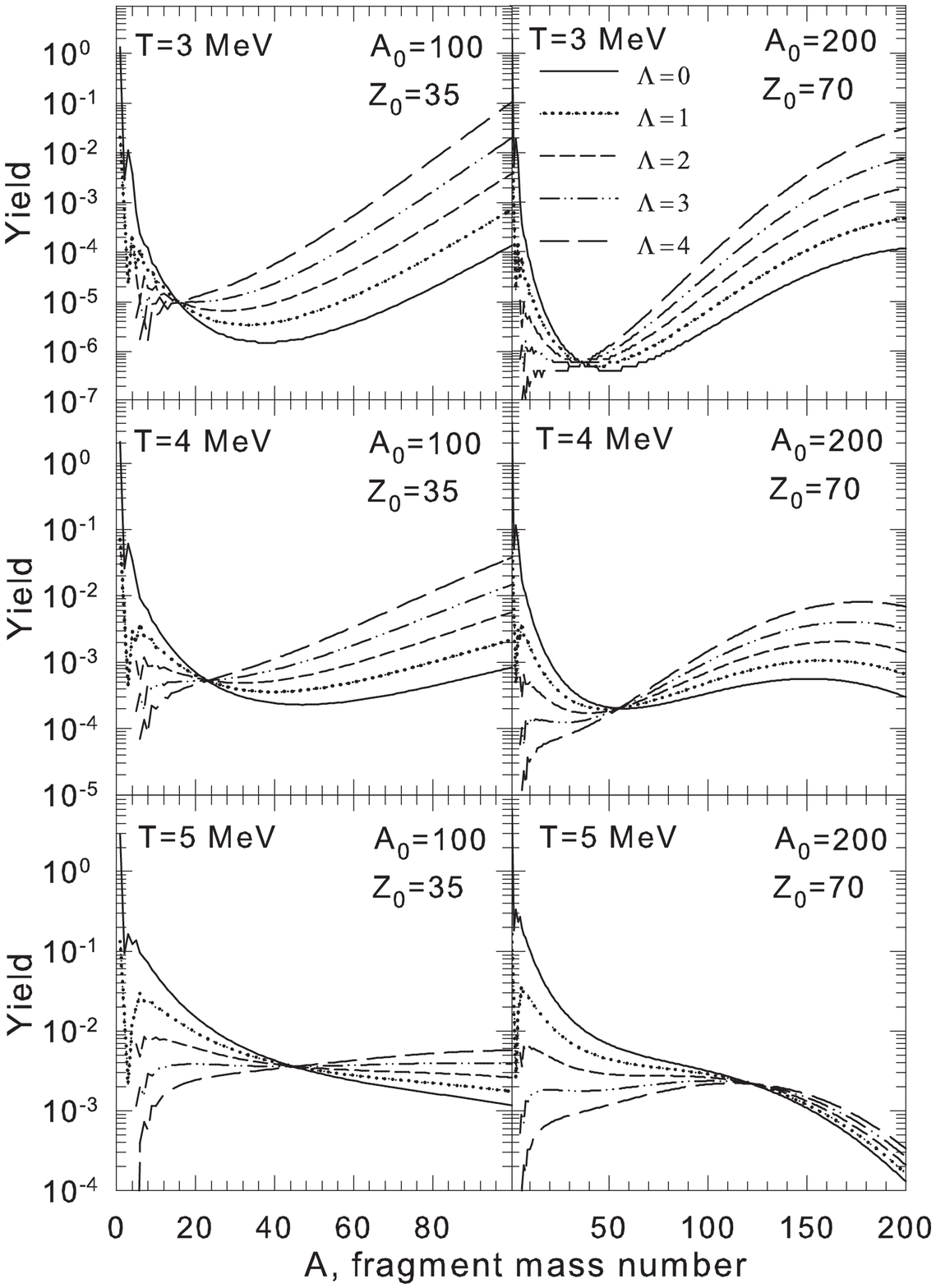}
\caption{\small{The same as Fig.~7, but for other neutron-rich systems. 
}}
\label{fig8}
\end{figure}

In all the figures one can see a very interesting feature: Curves of 
fragment yields with different numbers of $\Lambda$'s intersect 
at one point, at a particular $A$, which is determined by the 
parameters of the systems. Actually, this has a trivial explanation 
following the grand canonical structure of Eq.~(\ref{yazh}) and taking 
into account Eq.~(\ref{hyp}): For fragments with any hyperon content there 
is the same $A$ at which the sum contribution of $H$ into $Y_{\rm AZH}$ 
becomes 0. This happens when $(10.68 - 21.27/A^{1/3}) + \xi =0$. 
As shown in Ref.~\cite{bot-poch}, at very high temperatures $\xi$ 
becomes too low, the system disintegrates into small pieces only, and 
the mass distributions of all fragments decrease exponentially with $A$, 
without intersection. 

An important question for the examination of drip lines is the isotope 
composition of the produced fragments. For this purpose in Fig.~9 we 
present the isotope distributions of fluorine and molybdenum elements 
formed after disintegration of the systems with $A_0=$200 at the temperature 
$T=4$ MeV and with total charges $Z_0=$80, 70, and 60. Here we attempt to 
approach the neutron drip line by taking an excess of neutrons over protons. 
The first two cases give relatively moderate and large isospins in 
the system, which should be 
reachable with the FRS in future FAIR experiments. The last system is very 
neutron rich, however, this high isospin content may be considered as a 
limit that should be investigated to estimate the usefulness of the suggested 
method. Actually, in the case of relativistic projectiles, it is sufficient 
that such nuclear systems coming after the FRS would be rather short-lived 
(microseconds) 
in order to use them and generate new reactions leading to production 
of hypernuclei. As before we assumed that up to four $\Lambda$ hyperons were 
initially captured in the systems during the dynamical stage. 

\begin{figure}[tbh]
\includegraphics[width=0.8\textwidth]{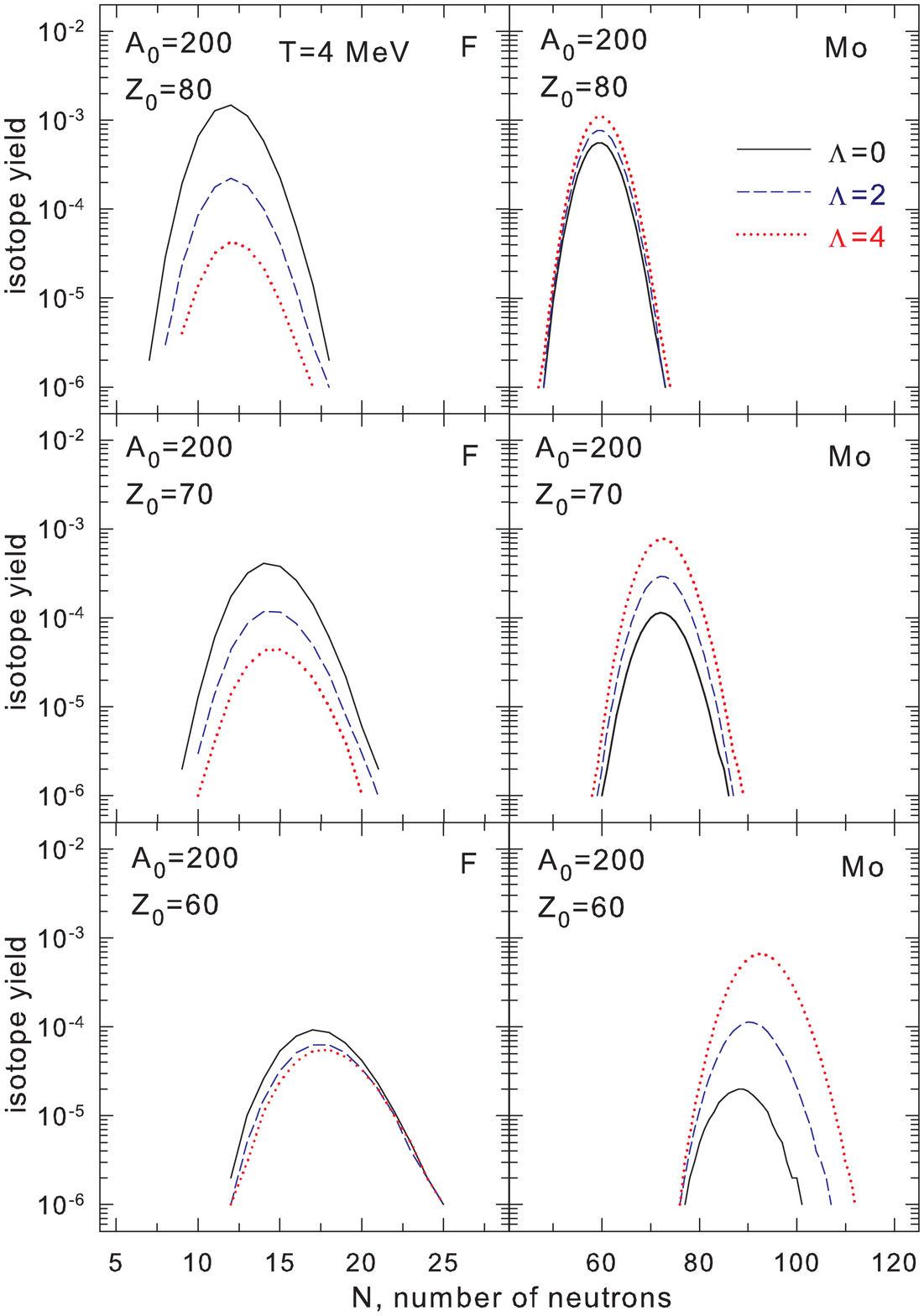}
\caption{\small{ (Color online)
Isotope yields of nuclei and hypernuclei of fluorine (F) and 
molybdenum (Mo) 
elements versus their neutron number (N) as calculated by the standard SMM for 
temperature $T=4$ MeV and for a system initially containing four $\Lambda$ 
hyperons. Initial mass numbers $A_0$ and charges $Z_0$ 
are shown. Lines are calculations for 
nuclei produced with a certain number of $\Lambda$ hyperons.
}}
\label{fig9}
\end{figure}

One can see in Fig.~9 that the absolute yields of isotopes change with the 
number of hyperons in fragments. This is related to the difference in 
mass yields discussed above. On the other hand, 
the isotope distribution widths look similar. 
The position of the peak is mainly determined by the system's isospin, 
however, the widths are quite large. For clarity, in the figure we show 
isotopes containing even numbers of hyperons only. The yields of isotopes 
with odd numbers of hyperons are in between the corresponding even numbers. 

We have also performed SMM calculations for systems with a 
smaller number of dynamically captured hyperons ($H_0=2$), which are more 
likely to be obtained in reactions. We have taken into account that the symmetry 
energy coefficient $\gamma$ in disintegration 
reactions may decrease from 25 to 14 MeV as discussed literature 
\cite{ogul}. In Fig.~10 we demonstrate a comparison between these calculations 
for the cases of two neutron-rich systems with large isospins 
($A_0$,$Z_0$) = (200,70) and (200,60). In addition, we show calculations for 
a proton-rich system with $A_0$=125 and $Z_0$=60, which is used to estimate 
possibilities of producing nuclei beyond the proton drip line.

\begin{figure}[tbh]
\includegraphics[width=0.8\textwidth]{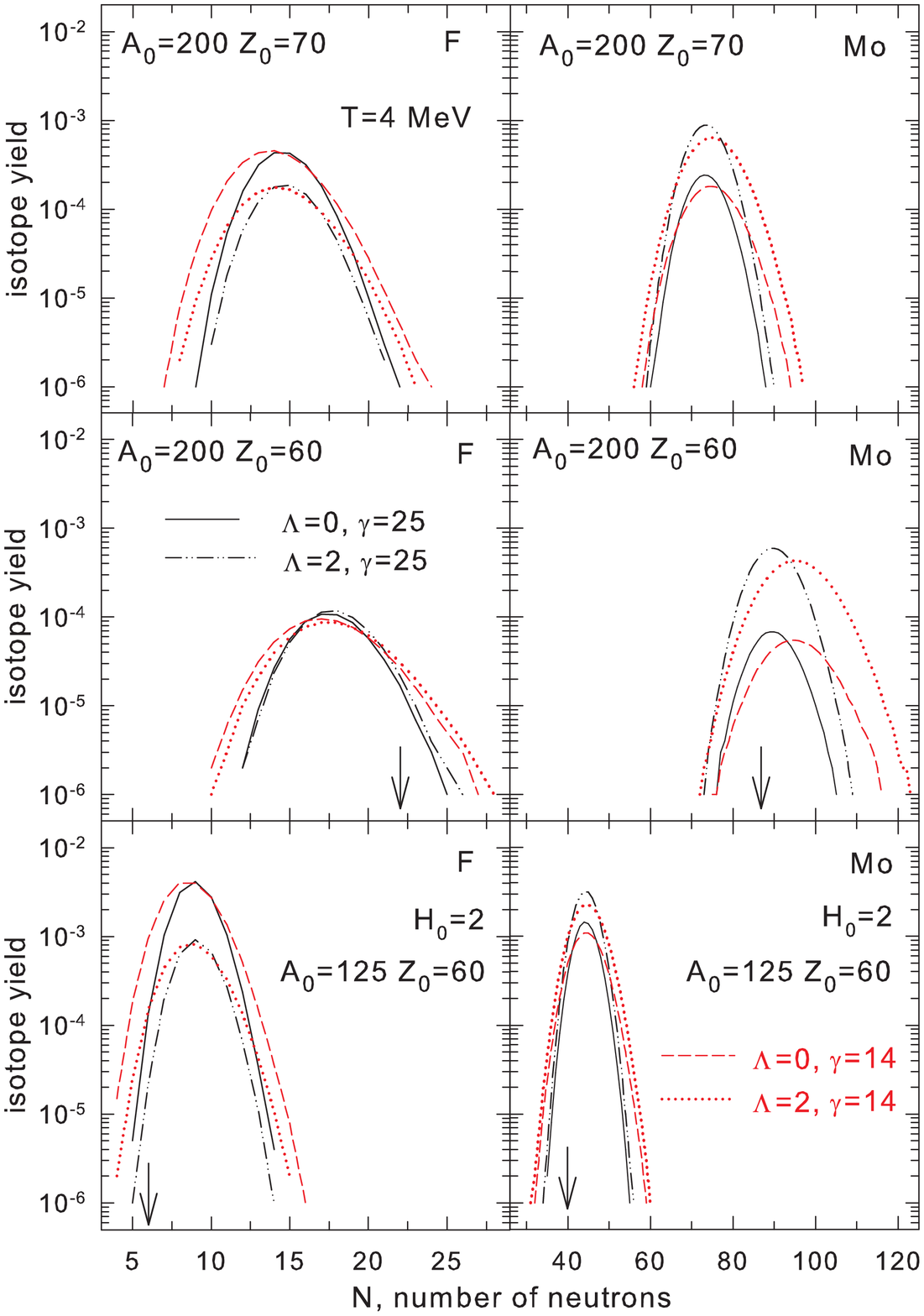}
\caption{\small{ (Color online)
The same as Fig.~9, but for systems initially containing $H_0=$2 
$\Lambda$ hyperons. Calculations with the standard symmetry energy 
coefficient $\gamma = 25$ MeV and the reduced symmetry energy $\gamma = 14$ 
MeV are presented. Arrows indicate approximate numbers 
of neutrons corresponding to the neutron and proton drip lines. 
}}
\label{fig10}
\end{figure}

It is known from normal nuclear multifragmentation that a decreasing 
symmetry energy leads to increasing yields of neutron-rich nuclei \cite{ogul}. 
In Fig.~10 we see that the distribution widths increase considerably 
at $\gamma=$14 MeV, and more neutron-rich isotopes can be produced, especially 
for large elements. For example, the expected neutron drip line for Mo with 
$N\approx 86-88$ can be reached and overpassed already after disintegration 
of a system 
with a moderately large isospin ($Z_0=70$). If it is possible to get 
secondary beams with even larger isospins, then, certainly, there will be 
abundant production of nuclei beyond the drip line. There is the same effect 
for fluorine too. Moreover, as shown in Fig.~10 the isotope distributions 
of normal nuclei (noted by $\Lambda$=0) 
behave similarly to those of hypernuclei. This can be explained simply by the same 
structure of the symmetry energy; see Eq.~(\ref{symmetry}). In this respect, 
using isospin-enriched relativistic beams for secondary interactions is a 
very promising method for study of normal nuclei too.
However, a big advantage of hyperfragments is that they are more stable, 
as the separation energies of nucleons are essentially higher. 
Therefore, neutrons that are accumulated in nuclei in the 
freeze-out volume will not leave them during the time scale of 
strong interaction when these nuclei leave the freeze-out. 
Possible secondary de-excitation processes will also affect these nuclei 
less, because of the high binding energy. 

Generally, the reaction mechanisms used in our approach suggest 
that proton-rich nuclei and proton-rich hypernuclei can be obtained in 
similar ways. This is a consequence of the regular behavior of neutron and proton 
separation energies (see Figs.~5 and 6). We have checked it by 
calculating just proton-rich systems. The general evolution of mass yields of 
fragments and hyperfragments with temperature is the same as shown 
in Figs.~7 and 8. The main conclusions regarding isotope production are also 
the same, as one can see in the bottom panels in Fig.~10. One can get proton-rich hyperfragments beyond the proton drip lines, which could also 
be stable with respect to strong decay. 

In the following these exotic proton- and neutron-rich $\Lambda$ hypernuclei 
will decay 
in weak processes. For heavy nuclei we expect predominantly nonmesonic 
decay producing two fast nucleons, like $\Lambda N \rightarrow NN$. 
With a considerable probability 
these nucleons can leave the nucleus without interactions, therefore, 
the large neutron/proton content of the nucleus will be preserved. This process 
gives us the opportunity, with the help of hypernuclei, to move beyond the 
drip lines in normal nuclei too and investigate islands of stability 
that may exist (e.g., \cite{greiner} and \cite{Yilmaz11}). This is 
a new method for obtaining such proton- and neutron-rich nuclei. 
The involvement of hyperons may provide a unique reaction mechanism to 
obtain nuclei with exotic isospins and to 
study both the proton and the neutron sides of the nuclear chart.


\section{Conclusion}

New promising reaction mechanisms for production of hypernuclei are 
under theoretical 
investigation. The dynamical stage of relativistic ion collisions can lead to the production of hyperons, which are captured by spectator residues. 
Disintegration of these hot residues leads to production of hypernuclei. 
This process was previously associated with the liquid-gas-type phase 
transition in finite nuclear systems and it provides the opportunity to study 
hypernuclear matter at subnuclear densities too. 
We demonstrate a broad variety of hypernuclei obtained in this way. 
The generalized SMM applied previously for description 
of disintegration processes in normal nuclei is also a good candidate to 
describe the hypernuclear case. We show that the nucleon separation 
energies in hypernuclei become considerably higher than in normal nuclei, 
because of coupling hyperons and nucleons inside nuclei. This makes it possible 
to obtain very exotic hypernuclei in these reactions, which can go far beyond 
the drip lines established for normal nuclei. Investigation of such 
hypernuclei will help to answer many fundamental questions of hyperphysics 
and nuclear physics. 
Moreover, the production of exotic hypernuclei in these reactions followed 
by their weak decay is a novel process which can be used to obtain normal exotic 
nuclei beyond the drip lines. It may provide a unique chance to investigate 
nuclear islands of stability. 
Presently, there are quite limited possibilities of 
obtaining such exotic hypernuclei with other experimental methods. 
We believe that hypernuclear and nuclear physics will benefit strongly from the exploriation of new production mechanisms for hypernuclei 
associated with spectator fragmentation reactions.

\begin{acknowledgments}
This work was supported by the the GSI Helmholtzzentrum f\"ur 
Schwerionenforschung and Hessian initiative for excellence 
(LOEWE) through the Helmholtz International Center for FAIR (HIC for FAIR), 
and by the Helmholtz-Institut Mainz. N.B. and A.S.B. 
thank the Frankfurt Institute for Advanced Studies and the Institut f{\"u}r 
Kernphysik of J.Gutenberg-Universit{\"a}t Mainz for hospitality. 
We also acknowledge the support from the Research Infrastructure Integrating 
Activity Study of Strongly Interacting Matter HadronPhysics3 under the Seventh 
Framework Programme of the EU (SPHERE network). 
\end{acknowledgments}

\end{document}